\documentclass[acmsmall,nonacm]{acmart}

\usepackage{pifont}
\usepackage{tikz}
\usetikzlibrary{arrows.meta,positioning}

\newcommand{\Instr}{\texttt{Instructor}}
\newcommand{\Proj}{\texttt{Project}}
\newcommand{\Stud}{\texttt{Student}}
\newcommand{\Pers}{\texttt{Person}}
\newcommand{\Dept}{\texttt{Department}}
\newcommand{\guide}{\texttt{guide}}
\newcommand{\work}{\texttt{work}}
\newcommand{\advise}{\texttt{advise}}

\newcommand{\Card}[3]{\mathrm{Card}(#1;\allowbreak\,#2;\allowbreak\,#3)}
\newcommand{\set}[1]{\{#1\}}
\newcommand{\cmp}[2]{\ensuremath{#1 \to #2}}   
\newcommand{\explicit}{\textit{explicit}}
\newcommand{\implied}{\textit{implied}}
\newcommand{\bydefault}{\textit{default}}

\makeatletter
\AtEndPreamble{%
  \@ifundefined{remark}{%
    \theoremstyle{remark}%
    \newtheorem{remark}[theorem]{Remark}%
    \theoremstyle{acmplain}%
  }{}}
\makeatother

\tikzset{
  ent/.style   = {circle, draw, minimum size=6.5mm, inner sep=0pt, font=\footnotesize},
  tup/.style   = {rectangle, draw, rounded corners=1.5pt, inner sep=3.5pt, font=\footnotesize},
  ar/.style    = {-{Latex[length=1.5mm,width=1.2mm]}, thin, draw=black!70},
  clulab/.style= {font=\footnotesize\itshape}
}

\title{Look Here or Look Across: Unified Cardinality Constraints for N-ary Relationships}

\author{Huanyi Chen}
\affiliation{%
  \institution{University of Waterloo}
  \department{PhD Alumnus}
  \country{Canada}}
\email{huanyi.chen@uwaterloo.ca}

\renewcommand{\shortauthors}{Chen}

\ccsdesc[500]{Information systems~Entity relationship models}
\ccsdesc[300]{Social and professional topics~Computing education}
\ccsdesc[100]{Mathematics of computing~Discrete mathematics}

\keywords{Entity--relationship model, cardinality constraints, n-ary relationships,
conceptual modelling, database design, computing education}

\begin{document}

\begin{abstract}
\noindent
A cardinality constraint written on an edge of an entity--relationship diagram
admits two opposite readings. Under the reading used by UML and by Chen's
original model, the label is read with the entity set on the \emph{same} side of
the relationship; under the reading used by standard database textbooks, it is
read with the entity set on the \emph{opposite} side. The two readings are exact
opposites, so a reader who assumes the wrong one takes away the opposite of what
the designer meant. For binary relationships the difficulty is confined to
interpretation, because the two constraints a binary relationship set admits can
both be drawn. For relationships among three or more entity sets the picture is
worse: a ternary relationship set admits twelve cardinality constraints, and a
diagram with three edges can carry at most three of them. This paper presents a
notation, $\Card{R}{p}{q} = (\mathit{lower}, \mathit{upper})$, that removes the
ambiguity without taking a side in it, and that is not limited to one constraint
per edge. We give the number of constraints an $n$-ary relationship set admits,
show which of them a design determines without ever writing them down, and state
two inference rules, \emph{decomposition} and \emph{augmentation}, that derive
one constraint from another, together with the side conditions under which each
is sound. We close with a worked case study that turns three business
requirements into three explicit constraints and nine more that the design
decides on its own.
\end{abstract}

\maketitle

\section{Introduction}

Entity--relationship (ER) modelling is one of the essential topics of a database
course. Its purpose is to equip students to turn business requirements into a
database schema, and the step where that translation is actually made is the
assignment of cardinality constraints \citep{chenEntityrelationshipModelUnified1976,
oliveConceptualModelingInformation2007}.

Correctly understanding those constraints is therefore critical, and it is also
where students most often go wrong. In our experience nine out of ten students
fail to interpret a cardinality constraint correctly, and the true proportion
may be worse. The reason is not that the concept is difficult. It is that there
are \emph{two} interpretations of a cardinality constraint in an ER diagram, and
that the two are \emph{exact opposites}. The two interpretations are scattered
without warning across textbooks, tools, and the web, so a student who reads a
diagram has no way of telling which one the designer had in mind.

Consider the familiar requirement set for a university database:

\begin{enumerate}
  \item A person can be assigned to at most two departments at a time.
  \item For each assigned department, a person cannot work on more than one
        project.
  \item Every project must be associated with at least one department, with at
        least one person working on it.
\end{enumerate}

\noindent
It is important that these constraints be enforced in the database, because they
become difficult to enforce once real data has been populated. But before they
can be enforced they must be written down, and it is precisely at that step that
the two readings diverge.

\paragraph{Contributions.}
\begin{itemize}
  \item We name the two readings, \emph{look here} and \emph{look
        across}, and show that they are not two theories but one object with its
        arguments exchanged (\S\ref{sec:two-readings}).
  \item We introduce the notation $\Card{R}{p}{q} = (\mathit{lower},
        \mathit{upper})$, which states a constraint without committing to either
        reading and without limiting a relationship set to one constraint per
        edge, and we give the number of constraints an $n$-ary relationship set
        admits: $3^n - 2^{n+1} + 1$ (\S\ref{sec:notation}).
  \item We show that a constraint nobody wrote down still has a meaning, that
        the default is $(0,*)$, and that a design determines many of the
        remaining constraints on its own (\S\ref{sec:implicit}).
  \item We state two inference rules, \emph{decomposition} and
        \emph{augmentation}, that derive one constraint from another, together
        with the side conditions that make each sound, and we define what it is
        for a constraint to be implied, redundant, or contradictory
        (\S\ref{sec:rules}).
  \item We work a case study in which three business requirements become three
        explicit constraints and nine further constraints that the design decides
        (\S\ref{sec:case-study}).
\end{itemize}

\section{Two Readings of a Cardinality Constraint}
\label{sec:two-readings}

\subsection{The binary case}

\begin{figure}[tbp]
  \centering
  \includegraphics[width=0.85\linewidth]{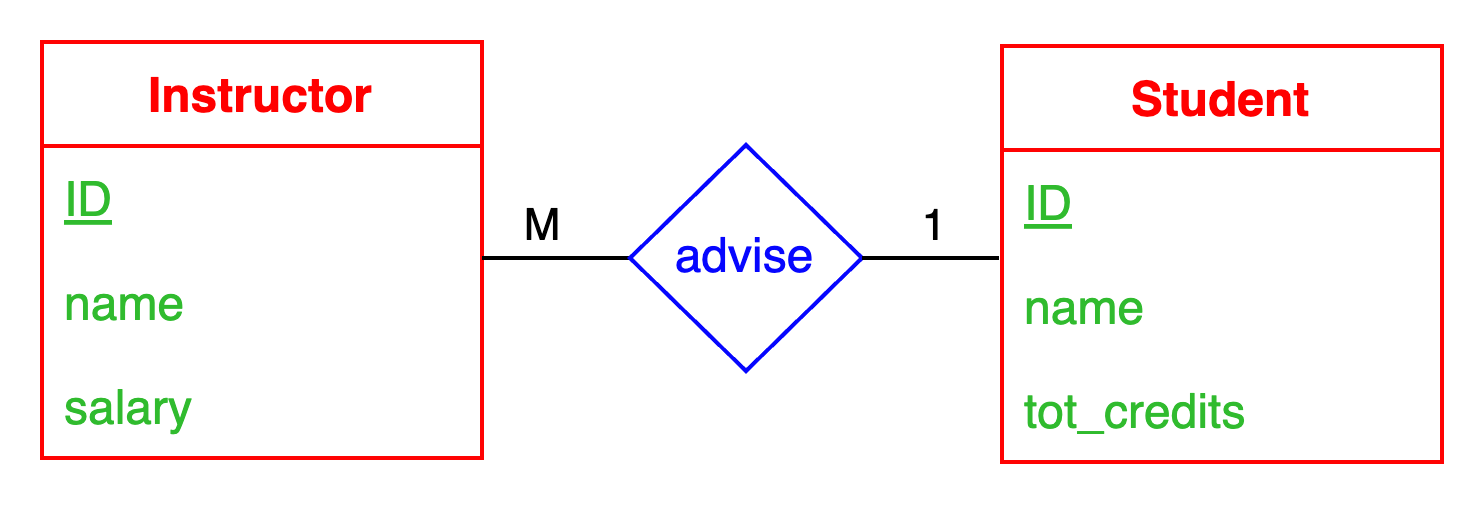}
  \caption{A binary relationship set carrying two cardinality labels. The $M$
           sits on the \Instr{} edge.}
  \Description{A binary relationship set: Instructor and Project joined through a diamond labelled advise, with a single label M on the Instructor edge.}
  \label{fig:advise-binary}
\end{figure}

Let us take two entity sets, \Instr{} and \Stud{}, and a relationship set
\advise{} between them. Suppose a designer draws the design of
Figure~\ref{fig:advise-binary}, in which $M$ indicates $1..*$ (many, and total
participation) and $1$ indicates $1..1$ (one, and total participation).

How should the design be interpreted? Is it many-to-one, or one-to-many?

The many-to-one reading is very intuitive, since from left to right the letters
can be read as ``$M$-to-$1$''. But the semantics of that reading are troubling.
Under it the diagram would say that many, indeed arbitrarily many, instructors
advise one student. That does not describe a university.

The other reading, one-to-many, is the one that survives, and the two readings
divide along the following lines.

\paragraph{UML and Chen's ER (look here).} UML records cardinality as a
multiplicity on each end of an association, and in UML's reading the design above
is \emph{many-to-one}. Chen's model, the original ER notation
\citep{chenEntityrelationshipModelUnified1976}, reads it the same way. Both are
intuitive once stated: the label $M$ is read along with the entity set on the
\emph{same} side as the label (here \Instr{}), so $M$ counts
\Instr{} instances. Read from a \Stud{}, there can be up to $M$ \Instr{}s.

\paragraph{The textbook's ER (look across).} In the ER model taught by
textbooks such as \citet{silberschatzDatabaseSystemConcepts2020}, the design
above is \emph{one-to-many}. The rationale comes from how ER diagrams are
eventually translated into relational schemas, and it is easier to see by
thinking of the translation directly:

\begin{itemize}
  \item each instructor can appear in \advise{} up to $M$ times;
  \item each student can appear in \advise{} only once.
\end{itemize}

\noindent
The resulting \texttt{Advise} relation has no duplication, because it is a
relational model:

\begin{center}
\begin{tabular}{cc}
\toprule
Instructor & Student \\
\midrule
Instructor 1 & Student 1 \\
Instructor 1 & Student 2 \\
Instructor 1 & Student 3 \\
$\vdots$     & $\vdots$     \\
\bottomrule
\end{tabular}
\end{center}

\noindent
This is the stage at which we check whether the cardinality constraint is
satisfied, and the relation plainly reflects a one-to-many relationship. Here the
label is interpreted relative to the entity set on the \emph{opposite} side,
\Stud{}, so $M$ counts \Stud{} instances.

\begin{figure}[tbp]
  \centering
  \includegraphics[width=0.85\linewidth]{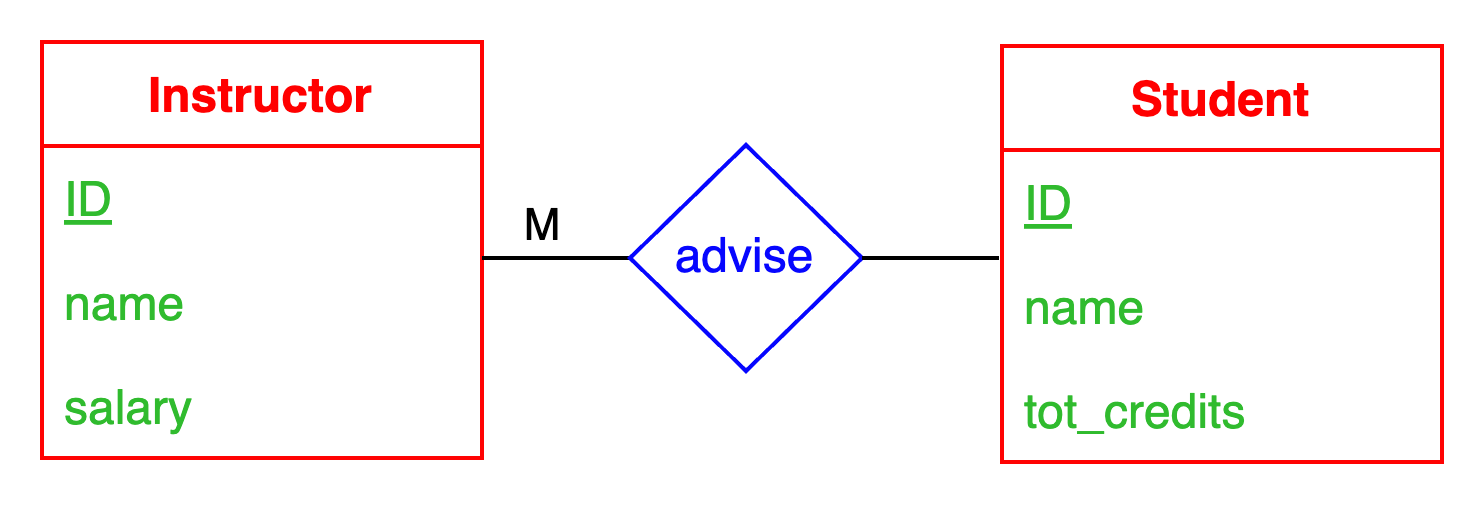}
  \caption{With only one cardinality constraint. The single $M$ is still read
           with the entity set on one side or the other.}
  \Description{Instructor and Project joined through an advise diamond, carrying only the single label M on the Instructor edge.}
  \label{fig:advise-only-m}
\end{figure}

\paragraph{The distinction survives with a single label.} It is tempting to
think that the disagreement needs two labels to show itself. It does not.
Figure~\ref{fig:advise-only-m} carries only the $M$ on the \Instr{} edge, and it
is still read two ways:

\begin{itemize}
  \item \textbf{UML or Chen's ER:} for each \emph{student}, there can be up to
        $M$ \emph{instructors};
  \item \textbf{the textbook's ER:} for each \emph{instructor}, there can be up
        to $M$ \emph{students}.
\end{itemize}

\noindent
This is what is meant by \emph{look here} and \emph{look across}. It is ``look
here'' in UML because the $M$ is read with the entity set on the same side,
\Instr{} in this case. It is ``look across'' in the textbook's ER because the $M$
is read with the entity set on the opposite side, \Stud{}.

\subsection{N-ary relationships}

Look here and look across are not specific to binary relationships. They apply
unchanged to relationships of any arity, and the disagreement does not shrink
with the arity. It grows.

\begin{figure}[tbp]
  \centering
  \includegraphics[width=0.55\linewidth]{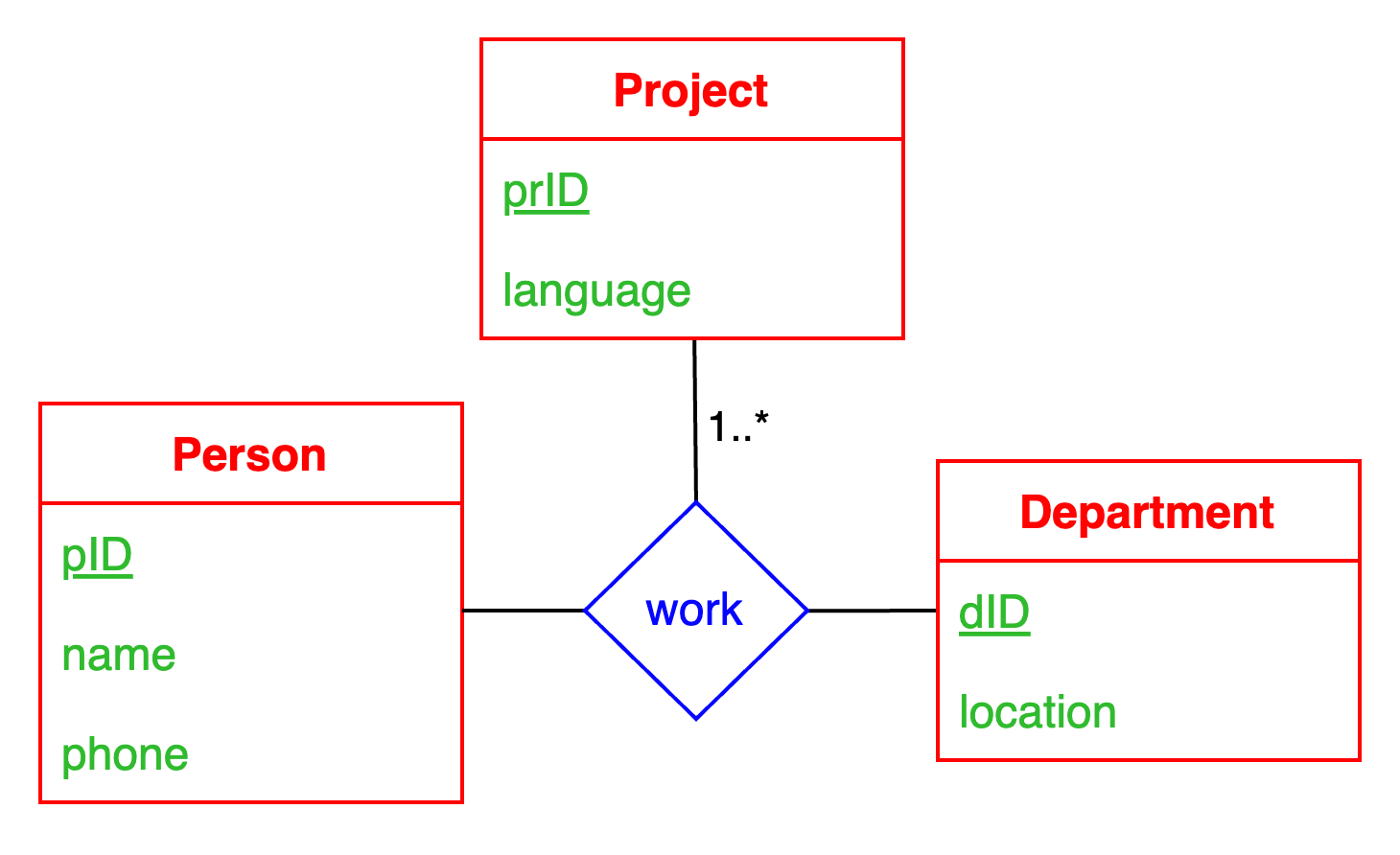}
  \caption{A ternary relationship set \work{} among \Pers{}, \Dept{}, and
           \Proj{}, carrying a single $1..*$ on the \Proj{} edge.}
  \Description{A ternary relationship set: Person, Department and Project joined through a diamond labelled work, with a single 1..* on the Project edge.}
  \label{fig:work-ternary}
\end{figure}

\noindent
Consider the ternary relationship set \work{} of Figure~\ref{fig:work-ternary},
among \Pers{}, \Dept{}, and \Proj{}, with the single label $1..*$ on the \Proj{}
edge. The two readings are:

\begin{itemize}
  \item \textbf{UML or Chen's ER:} for each $(\Pers, \Dept)$ pair, there can be
        many \Proj{}s;
  \item \textbf{the textbook's ER:} for each \Proj{}, there can be many
        $(\Pers, \Dept)$ pairs.
\end{itemize}

\noindent
The same picture and the same label, again the opposite of one another. And this
is where a second problem appears. The diagram has one edge per entity set, and
each edge holds exactly one label, so a ternary design can carry at most three
constraints. Section~\ref{sec:notation} shows that there are twelve of them to
carry.

\noindent
This design is a university research centre, and Section~\ref{sec:case-study}
returns to it with three business requirements to satisfy. Until then we use a
second ternary example, \guide{} among \Instr{}, \Proj{}, and \Stud{}, whose
entity sets have shorter names to write inside a constraint; the choice of
example is arbitrary, and everything below holds for either.

\section{A Unified Notation}
\label{sec:notation}

\subsection{How many cardinality constraints are there?}

Consider a ternary relationship set \guide{} among \Instr{}, \Proj{}, and
\Stud{}:

\begin{figure}[tbp]
  \centering
  \includegraphics[width=0.55\linewidth]{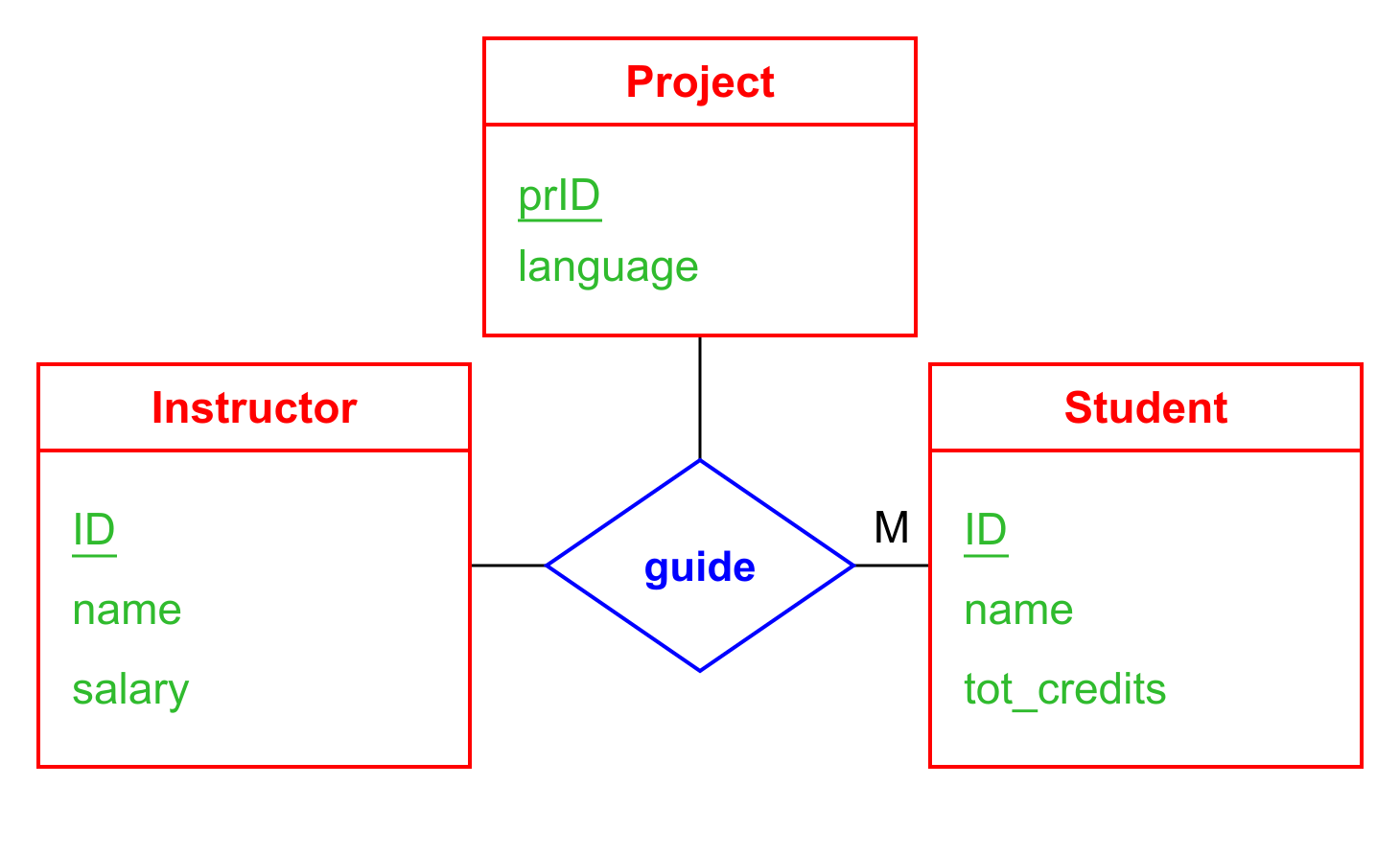}
  \caption{A ternary relationship set. The single $M$ sits on the \Stud{} edge.}
  \Description{A ternary relationship set: Instructor, Project and Student joined through a diamond labelled guide, with a single M on the Student edge.}
  \label{fig:guide}
\end{figure}

\noindent
A cardinality constraint says something of the form ``for each $X$, the number of
$Y$ is between \emph{lower} and \emph{upper}'', where $X$ and $Y$ are disjoint
groups of the entity sets participating in the relationship. With three entity
sets there is more than one way to choose those groups, so let us enumerate them.

Because every constraint below is on \guide{}, we write $\cmp{p}{q}$ as a
shorthand for $\Card{\guide}{p}{q}$. The twelve constraints come in \emph{six
pairs}: the right column of each row is the left column with $p$ and $q$
exchanged.

\begin{center}
\small
\begin{tabular}{ll}
\toprule
Constraint & Constraint (reversed) \\
\midrule
\cmp{\set{\Instr}}{\set{\Proj}}          & \cmp{\set{\Proj}}{\set{\Instr}} \\
\cmp{\set{\Instr}}{\set{\Stud}}          & \cmp{\set{\Stud}}{\set{\Instr}} \\
\cmp{\set{\Proj}}{\set{\Stud}}           & \cmp{\set{\Stud}}{\set{\Proj}} \\
\cmp{\set{\Instr,\Proj}}{\set{\Stud}}    & \cmp{\set{\Stud}}{\set{\Instr,\Proj}} \\
\cmp{\set{\Instr,\Stud}}{\set{\Proj}}    & \cmp{\set{\Proj}}{\set{\Instr,\Stud}} \\
\cmp{\set{\Proj,\Stud}}{\set{\Instr}}    & \cmp{\set{\Instr}}{\set{\Proj,\Stud}} \\
\bottomrule
\end{tabular}
\end{center}

\noindent
Where does the number twelve come from? A constraint has to name a non-empty
group $p$ to iterate over and a non-empty group $q$, disjoint from $p$, to count.
Choosing $p$ to be one entity set leaves $2^2 - 1 = 3$ choices for $q$ (the
subsets of the two remaining entity sets, minus the empty one), and there are
three such choices of $p$, for nine constraints. Choosing $p$ to be two entity
sets leaves $2^1 - 1 = 1$ choice for $q$ (the subsets of the single remaining
entity set, minus the empty one), and there are three such choices of $p$, for
three more. The group $p$ cannot be all three entity sets, because then nothing
is left to count. That gives $9 + 3 = 12$.

The same argument generalises. Every element of the $n$ entity sets is assigned
to $p$, to $q$, or to neither, which is $3^n$ assignments; subtract the $2^n$ in
which $p$ is empty and the $2^n$ in which $q$ is empty, and add back the single
assignment in which both are empty:

\begin{equation}
  \#\text{constraints}(n) \;=\; 3^n - 2^{n+1} + 1 .
  \label{eq:count}
\end{equation}

\noindent
For $n = 2$ this gives $2$, which is the pair of readings of
Section~\ref{sec:two-readings}; for $n = 3$ it gives $12$; for $n = 4$ it gives
$50$.

\subsection{The expressiveness gap}

Now compare Equation~\eqref{eq:count} with what a diagram can carry. A diagram
for an $n$-ary relationship set has $n$ edges, and each edge holds exactly one
label, for at most $n$ constraints. The gap is immediate and it widens quickly:

\begin{center}
\begin{tabular}{rrrr}
\toprule
$n$ & constraints $= 3^n - 2^{n+1} + 1$ & labels a diagram can carry & fraction expressible \\
\midrule
2 & 2  & 2 & $100\%$ \\
3 & 12 & 3 & $25\%$ \\
4 & 50 & 4 & $8\%$ \\
5 & 180 & 5 & $2.8\%$ \\
\bottomrule
\end{tabular}
\end{center}

\noindent
For binary relationships the ambiguity is the whole of the problem: the diagram
can express both constraints the relationship set admits, but it cannot say which
of the two a given edge's label denotes. From three entity sets onward the
diagram is not merely ambiguous, it is inexpressive: most of the constraints a
designer cares about have nowhere to go.

Written out for the ternary case, the twelve constraints a designer might want
are these, and the diagram of Figure~\ref{fig:guide} can show at most three of
them:

\begin{center}
\begin{tabular}{rll}
\toprule
\# & Constraint & \\
\midrule
1  & \cmp{\set{\Instr}}{\set{\Proj}}          & \\
2  & \cmp{\set{\Instr}}{\set{\Stud}}          & \\
3  & \cmp{\set{\Proj}}{\set{\Stud}}           & \\
4  & \cmp{\set{\Instr,\Proj}}{\set{\Stud}}    & \\
5  & \cmp{\set{\Instr,\Stud}}{\set{\Proj}}    & \\
6  & \cmp{\set{\Proj,\Stud}}{\set{\Instr}}    & \\
7  & \cmp{\set{\Proj}}{\set{\Instr}}          & \\
8  & \cmp{\set{\Stud}}{\set{\Instr}}          & \\
9  & \cmp{\set{\Stud}}{\set{\Proj}}           & \\
10 & \cmp{\set{\Stud}}{\set{\Instr,\Proj}}    & \\
11 & \cmp{\set{\Proj}}{\set{\Instr,\Stud}}    & \\
12 & \cmp{\set{\Instr}}{\set{\Proj,\Stud}}    & \\
\bottomrule
\end{tabular}
\end{center}

\noindent
\noindent
{\setlength{\abovedisplayskip}{4pt plus 2pt minus 2pt}%
 \setlength{\belowdisplayskip}{4pt plus 2pt minus 2pt}%
 \setlength{\abovedisplayshortskip}{2pt plus 1pt minus 1pt}%
 \setlength{\belowdisplayshortskip}{2pt plus 1pt minus 1pt}%
 Write $M$ on the \Stud{} edge, and UML reads it as
 \[
   \Card{\guide}{\set{\Instr,\Proj}}{\set{\Stud}} = (1, M),
 \]
 while the textbook reads the same $M$ as
 \[
   \Card{\guide}{\set{\Stud}}{\set{\Instr,\Proj}} = (1, M).
 \]
 You cannot express both at once, and a reader who assumes the other convention
 takes away the opposite of what you meant.}

\subsection{Three ways to deal with cardinality constraints in an n-ary relationship}

As far as we know there are three ways out of this:

\begin{enumerate}
  \item Split the $n$-ary relationship into multiple binary relationships
        \ding{55}
  \item Add an artificial entity set \ding{55}
  \item Represent the cardinality constraints non-graphically \ding{51}
\end{enumerate}

\noindent
The first two are the usual workarounds, and both cost more than they look like
they do. Splitting a ternary relationship into binary ones changes what the
design means: three binary relationship sets can record three pairwise facts
without ever recording that the three belong together, and the decomposition is
lossless only under conditions that have to be checked
\citep{jonesTernaryRelationshipDecomposition1996}. An artificial entity set has
to be given an identity that the domain does not provide (its key is something
we invent rather than something we discover), and it only pushes the problem one
level down.

The third one is what the rest of this paper is about.

\subsection{Generalized cardinality constraints}

Instead of a mark on an edge, write the constraint as a function:

\begin{definition}[Generalized cardinality constraint]
\label{def:card}
$\Card{R}{p}{q} = (\mathit{lower}, \mathit{upper})$ states that for each $p$,
the number of tuples from $q$ is between $\mathit{lower}$ and
$\mathit{upper}$, where $R$ is the relationship set and $p$ and $q$ are disjoint
sets of entity sets.
\end{definition}

For the ternary example above, the two conventions become two lines:

\begin{center}
\begin{tabular}{ll}
\toprule
Convention & Constraint \\
\midrule
UML or Chen's ER (\emph{look here})   & \cmp{\set{\Instr,\Proj}}{\set{\Stud}} $= (1, M)$ \\
The textbook's ER (\emph{look across}) & \cmp{\set{\Stud}}{\set{\Instr,\Proj}} $= (1, M)$ \\
\bottomrule
\end{tabular}
\end{center}

\begin{remark}
Under Definition~\ref{def:card}, the two readings differ \emph{only} in which
side is named as $p$. Same picture,
same numbers, exchanged arguments. This is the whole disagreement from
Section~\ref{sec:two-readings}: there, the many-to-one reading was
$\Card{\advise}{\set{\Stud}}{\set{\Instr}} = (1, M)$ and the one-to-many reading
was $\Card{\advise}{\set{\Instr}}{\set{\Stud}} = (1, M)$, again the same two
numbers, with $p$ and $q$ the other way around.
\end{remark}

That is why we call this notation \emph{unified}. It does not force you to pick a
side of the look-here/look-across debate, and it does not care whether the
relationship set is binary or $n$-ary. Since every constraint names its own $p$
and $q$, there is nothing left to guess: you write the reading you mean, and a
reader, or a program, gets exactly that one.

\subsection{Encoding generalized cardinality constraints}

Since $\Card{R}{p}{q}$ is now a mathematical object rather than a mark on a
picture, it can be written down directly. Because the bounds are numbers, an
unbounded $M$ has to be replaced by a concrete value; where a design names a
specific $M$, we take $M = 10$ throughout, and we write $*$ for an upper bound
that is genuinely unbounded. The two conventions of the previous section then
become two lines, which can sit next to each other without contradiction because
each one names its own $p$ and $q$:

\begin{itemize}
  \item $\Card{\guide}{\set{\Instr,\Proj}}{\set{\Stud}} = (1, 10)$
  \item $\Card{\guide}{\set{\Stud}}{\set{\Instr,\Proj}} = (1, 10)$
\end{itemize}

\noindent
The real payoff is that we are no longer limited to one constraint per edge. The
same relationship set can carry four constraints at once:

\begin{itemize}
  \item $\Card{\guide}{\set{\Stud}}{\set{\Instr,\Proj}} = (1, 10)$
  \item $\Card{\guide}{\set{\Instr}}{\set{\Proj}} = (0, 2)$
  \item $\Card{\guide}{\set{\Instr}}{\set{\Stud}} = (1, 5)$
  \item $\Card{\guide}{\set{\Stud}}{\set{\Proj}} = (1, *)$
\end{itemize}

\noindent
Read out, this says that each student is guided on $1$ to $10$
$(\Instr, \Proj)$ combinations, that each instructor works on at most two
projects, that each instructor is associated with $1$ to $5$ students, and that
each student works on at least one project. Three of those four constraints have
no place on the diagram at all, and the fourth is the one whose reading you would
otherwise have had to guess. Note also that these constraints do not all follow
the same convention, and that this no longer matters: each one names its own $p$
and $q$, so it says what it means on its own.

\section{Implicit Constraints and the Default}
\label{sec:implicit}

Writing the constraints down as data is also what makes them something a program
can read, and once a program can read them, it can verify them. That raises the
question this section answers: given the explicit constraints, which implicit
cardinality constraints can exist, and which of those the design has already
decided?

\subsection{The constraint you never wrote}

Let us start with what a constraint means when it is not there.

\begin{definition}[Default]
If a cardinality constraint is not explicitly stated, it is
$\Card{R}{p}{q} = (0, *)$: for each $p$, the number of $q$s is somewhere between
zero and as many as there happen to be. In other words, nothing is being said.
\end{definition}

That is where the trouble starts. $(0, *)$ is what the design says when no
constraint was written, even when there should have been one.

\subsection{Twelve constraints, one of them written}

Return to the ternary relationship set \guide{} of Figure~\ref{fig:guide}, and
suppose the designer states exactly one of the twelve constraints:

\begin{center}
\begin{tabular}{rll}
\toprule
\# & Constraint & Bound \\
\midrule
1  & \cmp{\set{\Instr}}{\set{\Proj}}          & \\
2  & \cmp{\set{\Instr}}{\set{\Stud}}          & \\
3  & \cmp{\set{\Proj}}{\set{\Stud}}           & \\
4  & \cmp{\set{\Instr,\Proj}}{\set{\Stud}}    & \\
5  & \cmp{\set{\Instr,\Stud}}{\set{\Proj}}    & \\
6  & \cmp{\set{\Proj,\Stud}}{\set{\Instr}}    & \\
7  & \cmp{\set{\Proj}}{\set{\Instr}}          & \\
8  & \cmp{\set{\Stud}}{\set{\Instr}}          & \\
9  & \cmp{\set{\Stud}}{\set{\Proj}}           & \\
10 & \cmp{\set{\Stud}}{\set{\Instr,\Proj}}    & $(2,2)$ given \\
11 & \cmp{\set{\Proj}}{\set{\Instr,\Stud}}    & \\
12 & \cmp{\set{\Instr}}{\set{\Proj,\Stud}}    & \\
\bottomrule
\end{tabular}
\end{center}

\noindent
Constraint 10 says each student is guided on exactly two
$(\Instr, \Proj)$ combinations. What are the other eleven?

\subsection{Four constraints the design decides}

Take constraint 8, \cmp{\set{\Stud}}{\set{\Instr}}: for each student, how many
instructors?

Nothing was written about instructors, so the default says $(0, *)$. But the
constraint that \emph{was} written says each student has exactly two
$(\Instr,\Proj)$ pairs, and each of those pairs names one instructor. Two pairs
name at most two instructors, so the count is never three or more. It is never
zero either: constraint 10 puts its lower bound at $2$, so every student is in a
pair, and a student who is in a pair has an instructor. What is left is one or
two.

\begin{figure}[b]
\centering
\begin{minipage}[c]{0.47\linewidth}\centering
  {\footnotesize\itshape two instructors}\\[1.5mm]
  \begin{tikzpicture}[x=1mm,y=1mm]
    \node[ent] (g)  at ( 0, 0)    {$s_1$};
    \node[tup] (t1) at (17, 4.5)  {$(i_1,p_1,s_1)$};
    \node[tup] (t2) at (17,-4.5)  {$(i_2,p_2,s_1)$};
    \node[ent] (c1) at (34, 4.5)  {$i_1$};
    \node[ent] (c2) at (34,-4.5)  {$i_2$};
    \draw[ar] (g)--(t1);  \draw[ar] (g)--(t2);
    \draw[ar] (t1)--(c1); \draw[ar] (t2)--(c2);
  \end{tikzpicture}
\end{minipage}\hfill
\begin{minipage}[c]{0.47\linewidth}\centering
  {\footnotesize\itshape one instructor}\\[1.5mm]
  \begin{tikzpicture}[x=1mm,y=1mm]
    \node[ent] (g)  at ( 0, 0)    {$s_1$};
    \node[tup] (t1) at (17, 4.5)  {$(i_1,p_1,s_1)$};
    \node[tup] (t2) at (17,-4.5)  {$(i_1,p_2,s_1)$};
    \node[ent] (c1) at (34, 0)    {$i_1$};
    \draw[ar] (g)--(t1);  \draw[ar] (g)--(t2);
    \draw[ar] (t1)--(c1); \draw[ar] (t2)--(c1);
  \end{tikzpicture}
\end{minipage}
\caption{Why constraint 8 is $(1,2)$. The node on the left is the group the
         constraint iterates over, the boxes are the tuples of \guide{} that
         group appears in, and the circles are the entities being counted. The
         count is the number of circles, not the number of arrows. Each diagram
         reads the same way throughout this paper.}
\Description{Two graphs side by side. Each has a Student node on the left, tuple boxes in the middle, and Instructor nodes on the right; the first shows two instructors for one student, the second a single instructor in both tuples.}
\label{fig:two-one}
\end{figure}
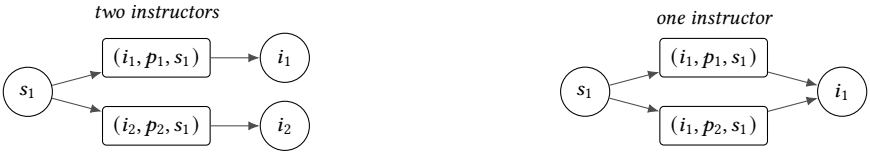

\noindent
So we get $(1, 2)$ for constraint 8, and neither end of the interval can be
moved: $1$ happens when both pairs name the same instructor, and $2$ happens when
they name different ones. The two clusters of Figure~\ref{fig:two-one} are two
different instances, each satisfying constraint 10, one for each end. A student
has one count or the other, and the design does not say which.

Constraint 9, \cmp{\set{\Stud}}{\set{\Proj}}, is the same argument with the two
counted positions swapped, and is also $(1, 2)$.

Constraint 5, \cmp{\set{\Instr,\Stud}}{\set{\Proj}}, is decided the same way,
one level up. Given an instructor and a student together, how many projects? It
can be zero, since nothing requires a particular instructor and student to have a
project between them; and it is at most two, because a student has only two pairs
in total. Constraint 6, \cmp{\set{\Proj,\Stud}}{\set{\Instr}}, is the same
argument with the counted positions swapped, and is $(0, 2)$ with it.

That is four of the eleven, and none of them was in the document. This is the
ordinary case, not a curiosity: a constraint that follows from the ones we wrote
is not a second constraint to be drawn and maintained; it is a sentence a program
can produce on demand.

\subsection{Seven constraints the design does not decide}

The other seven come out as $(0, *)$. Here the default is not a placeholder for
an answer nobody worked out: it is the best possible answer, and the figure
below is the reason. Showing that no tighter bound follows takes a design that
reaches the bound, and $*$ is the harder of the two ends: what has to be reached
is a count that keeps growing.

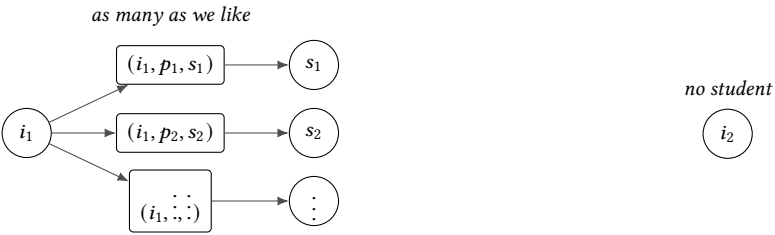
\begin{figure}[b]
\centering
\begin{minipage}[c]{0.47\linewidth}\centering
  {\footnotesize\itshape as many as we like}\\[1.5mm]
  \begin{tikzpicture}[x=1mm,y=1mm]
    \node[ent] (g)  at ( 0, 0)    {$i_1$};
    \node[tup] (t1) at (19, 9)    {$(i_1,p_1,s_1)$};
    \node[tup] (t2) at (19, 0)    {$(i_1,p_2,s_2)$};
    \node[tup] (t3) at (19,-9)    {$(i_1,\vdots,\vdots)$};
    \node[ent] (c1) at (38, 9)    {$s_1$};
    \node[ent] (c2) at (38, 0)    {$s_2$};
    \node[ent] (c3) at (38,-9)    {$\vdots$};
    \draw[ar] (g)--(t1); \draw[ar] (g)--(t2); \draw[ar] (g)--(t3);
    \draw[ar] (t1)--(c1); \draw[ar] (t2)--(c2); \draw[ar] (t3)--(c3);
  \end{tikzpicture}
\end{minipage}\hfill
\begin{minipage}[c]{0.47\linewidth}\centering
  {\footnotesize\itshape no student}\\[1.5mm]
  \begin{tikzpicture}[x=1mm,y=1mm]
    \node[ent] (g) at (0,0) {$i_2$};
  \end{tikzpicture}
\end{minipage}
\caption{The two ends of $(0,*)$. The left cluster grows without bound; the
         right one has a group that appears in no tuple at all and so counts
         zero. Only the tuples that carry the count are drawn: a student's
         second pair is left out, and putting it back with a fresh instructor
         and project keeps constraint 10 satisfied without changing any of
         these counts.}
\Description{Two graphs side by side. The left shows one Instructor joined through several tuple boxes to an unbounded number of Students; the right shows an Instructor node with no tuples at all.}
\label{fig:zero-star}
\end{figure}

\noindent
\textbf{Constraints 1, 2, 3, and 7} fix one entity and count entities of another
type. One instructor can be in as many pairs as there are students in the
relationship set, so constraint 2 has no upper bound: constraint 10 says each
student has two pairs, and says nothing about how many students share an
instructor. Give one instructor a different project for each of their students,
and their project count grows with the number of students, which is constraint 1;
the same move gives constraint 7 a project with as many instructors as there are
students to work on it, and constraint 3 a project shared by as many students.
Figure~\ref{fig:zero-star} is that picture.

\medskip
\noindent
\textbf{Constraints 4, 11, and 12} count a pair rather than an entity. Constraint
12 fixes an instructor and counts $(\Proj,\Stud)$ pairs: every tuple the
instructor is in contributes a different one, so the count grows with the
instructor's tuples. Constraint 11 is the same picture with the roles swapped, a
project and its $(\Instr,\Stud)$ pairs; constraint 4 fixes a pair instead of an
entity, which is the argument of constraint 5 above, with the same growth.

The lower bound of $0$ takes care of itself: a group that appears in no tuple at
all has a count of $0$, which is the right-hand cluster in the figure. Putting
the two halves together, we get:

\begin{center}
\begin{tabular}{rlll}
\toprule
\# & Constraint & Bound & \\
\midrule
1  & \cmp{\set{\Instr}}{\set{\Proj}}          & $(0,*)$ & \bydefault \\
2  & \cmp{\set{\Instr}}{\set{\Stud}}          & $(0,*)$ & \bydefault \\
3  & \cmp{\set{\Proj}}{\set{\Stud}}           & $(0,*)$ & \bydefault \\
4  & \cmp{\set{\Instr,\Proj}}{\set{\Stud}}    & $(0,*)$ & \bydefault \\
5  & \cmp{\set{\Instr,\Stud}}{\set{\Proj}}    & $(0,2)$ & \implied \\
6  & \cmp{\set{\Proj,\Stud}}{\set{\Instr}}    & $(0,2)$ & \implied \\
7  & \cmp{\set{\Proj}}{\set{\Instr}}          & $(0,*)$ & \bydefault \\
8  & \cmp{\set{\Stud}}{\set{\Instr}}          & $(1,2)$ & \implied \\
9  & \cmp{\set{\Stud}}{\set{\Proj}}           & $(1,2)$ & \implied \\
10 & \cmp{\set{\Stud}}{\set{\Instr,\Proj}}    & $(2,2)$ & given \\
11 & \cmp{\set{\Proj}}{\set{\Instr,\Stud}}    & $(0,*)$ & \bydefault \\
12 & \cmp{\set{\Instr}}{\set{\Proj,\Stud}}    & $(0,*)$ & \bydefault \\
\bottomrule
\end{tabular}
\end{center}

\noindent
One constraint given, four implied, seven settled by the default. The two rules
that produce the four, and what it means for one bound to follow from another,
are the subject of the next section.

\section{Inference Rules}
\label{sec:rules}

The previous section derived four constraints from one without stating a rule.
This section states the rules.

\subsection{Bounds are intervals}

A constraint $\Card{R}{p}{q} = (l, u)$ is contained in another, $(l', u')$-shaped
one, exactly when $l \geq l'$ and $u \leq u'$, that is, when its interval sits
inside the other. The contained constraint is the stronger statement, since it
rules out more instances. Two constraints on the same $(R; p; q)$ combine to
$(\max(l, l'), \min(u, u'))$, and the pair is contradictory exactly when that
interval is empty. Write $C_{\min}(R;p;q)$ and $C_{\max}(R;p;q)$ for the lower
and upper bound of $\Card{R}{p}{q}$, so that $\Card{R}{p;q}$ is the interval
$(C_{\min}(R;p;q), C_{\max}(R;p;q))$.

With that in hand:

\begin{itemize}
  \item A constraint is \textbf{implied} when the rest of the design already
        forces something at least as strong.
  \item It is \textbf{redundant} when it is implied and adds nothing.
  \item The design is \textbf{contradictory} when two written constraints have
        no interval in common.
\end{itemize}

\noindent
Two bounds that merely differ are neither. A design that says $(1,5)$ for
constraint 8, when constraint 10 forces $(1,2)$, is not in conflict with itself;
it just has a loose bound in it, which is the sort of thing a marker would want to
know and a program can point out for free.

\subsection{Decomposition}

Decomposition relates a group to a larger group that contains it. Each $q$
appears in at least one $(q,s)$ pair, so a $p$ can never have more $q$s than it
has $(q,s)$ pairs:

\begin{equation}
\begin{aligned}
  C_{\min}(R; p; q \cup s) &\geq C_{\min}(R; p; q), \\
  C_{\max}(R; p; q \cup s) &\geq C_{\max}(R; p; q).
\end{aligned}
\label{eq:decomposition}
\end{equation}

\noindent
Knowing a bound on $p$ over the pair-group $q \cup s$, we can therefore bound $p$
over the smaller group $q$. Constraint 10 is a bound on \Stud{} over
$\set{\Instr,\Proj}$, so decomposing gives bounds on \Stud{} over \Instr{} and
over \Proj{}: constraints 8 and 9, $(1,2)$ each.
Remark~\ref{rem:decomposition} separates what the rule transfers from what it does
not.

\begin{remark}[What decomposition transfers, and what it does not]
\label{rem:decomposition}
The second inequation of \eqref{eq:decomposition} is the one that carries a
number across: read right to left it says $C_{\max}(R;p;q) \leq
C_{\max}(R;p;q\cup s)$, so an upper bound on the pair-group is also an upper bound
on the subgroup. That is where the $2$ in constraints 8 and 9 comes from.

The first inequation transfers no upper bound on the minimum; it only says the
subgroup's minimum cannot exceed the pair-group's. The lower bound of $1$ in
constraints 8 and 9 therefore needs a separate argument, and the argument is a
counting one: constraint 10 forces every student into at least two pairs, and two
pairs name at least one instructor, so the count is at least $\lceil 2/2 \rceil =
1$. In general, if a $p$ has at least $m$ pairs and no $q$ can carry more than
$k$ of them, then it has at least $\lceil m/k \rceil$ distinct $q$s.
\end{remark}

\subsection{Augmentation}

Augmentation relates a group to a larger group that contains it on the other
side. An $(p,s)$ pair determines a $p$, and the $q$s available to the pair are
among those available to that $p$: every tuple that names the pair also names the
$p$, so the pair's count is at most the $p$'s. Both bounds therefore move
together, and the direction is downward:

\begin{equation}
\begin{aligned}
  C_{\min}(R; p \cup s; q) &\leq C_{\min}(R; p; q), \\
  C_{\max}(R; p \cup s; q) &\leq C_{\max}(R; p; q).
\end{aligned}
\label{eq:augmentation}
\end{equation}

\noindent
An upper bound carries over to the finer group; a lower bound does not, because a
pair may have fewer $q$s than the $p$ it refines. That asymmetry is what separates
constraints 5 and 6 from constraints 8 and 9. Constraint 10 caps \Stud{} at two
$(\Instr,\Proj)$ pairs, so decomposing it caps \Stud{} at two \Proj{}s and at two
\Instr{}s (constraints 9 and 8), and the counting argument of
Remark~\ref{rem:decomposition} supplies their lower bound, making each $(1,2)$.
Augmenting constraint 9 to the pair $\set{\Instr,\Stud}$, and constraint 8 to
$\set{\Proj,\Stud}$, caps those pairs at two as well: constraints 5 and 6,
$(0,2)$ each. The upper bound of $2$ is inherited; the lower bound of $0$ is not,
because nothing requires a given instructor and student to be working on anything
together.

\begin{remark}[The support condition]
\label{rem:support}
An upper bound carries from $p$ to $(p,s)$ unconditionally, since the pair's $q$s
are among the $p$'s. A lower bound does not travel with it.
\eqref{eq:augmentation} is stated over every pair, so a pair that appears in no
tuple contributes $0$ to $C_{\min}(R; p \cup s; q)$, and that is what keeps the
lower bound of constraints 5 and 6 at the default of $0$. A pair that does appear
has at least one $q$, so its own count is at least $1$. But that is a statement
about one pair, not about the group being iterated over.

For the rest of the paper a constraint is read as holding over \emph{every} group
$p$ that can be formed from the entity sets it names, whether or not that group
appears in a tuple. A group in no tuple contributes $0$ to the minimum, and that
is what makes $(0,*)$ the default rather than $(*,*)$. A constraint whose own
lower bound is $1$ or more forces every group it iterates over to appear, which is
the whole reason constraints 8 and 9 have a lower bound of $1$ while constraints 5
and 6 do not.
\end{remark}

\noindent
The transfer does not run the other way. Suppose the bound sits on the pair-group
instead: $\Card{\guide}{\set{\Instr,\Proj}}{\set{\Stud}} = (0,1)$, at most one
student per pair. That says nothing about how many students one instructor has.
Figure~\ref{fig:zero-star} shows both ends: one cluster gives $i_1$ a student per
pair, without limit, and the other gives $i_2$ none, since $i_2$ is in no pair.
The coarser constraint stays where it started:

\begin{center}
$\begin{aligned}
  &\Card{\guide}{\set{\Instr,\Proj}}{\set{\Stud}} = (0,1)\\
  \Longrightarrow\;&\Card{\guide}{\set{\Instr}}{\set{\Stud}} = (0,*)
\end{aligned}$
\end{center}

\noindent
The two rules lean in opposite directions and both lose information. Decomposition
shrinks the group being counted; augmentation refines the group being iterated
over. Neither ever produces a bound tighter than the default on its own
authority: the minimum of a derived constraint is $1$ only when the constraint it
came from forces it to be.

\subsection{A variant: one tuple per group}

Let us change the one constraint that is written, and the whole set of bounds
changes with it.

Suppose constraint 10 is $(1,1)$ instead of $(2,2)$: each student is guided on
exactly one $(\Instr,\Proj)$ pair. Then constraints 5 and 6 become $(0,1)$ (at
most one, with no lower bound), and constraints 8 and 9 become $(1,1)$, since a
student in exactly one pair has exactly one instructor and exactly one project.
The rest stay at the default.

\begin{center}
\begin{tabular}{rll}
\toprule
\# & Constraint & Bound \\
\midrule
5  & \cmp{\set{\Instr,\Stud}}{\set{\Proj}}    & $(0,1)$ \implied \\
6  & \cmp{\set{\Proj,\Stud}}{\set{\Instr}}    & $(0,1)$ \implied \\
8  & \cmp{\set{\Stud}}{\set{\Instr}}          & $(1,1)$ \implied \\
9  & \cmp{\set{\Stud}}{\set{\Proj}}           & $(1,1)$ \implied \\
10 & \cmp{\set{\Stud}}{\set{\Instr,\Proj}}    & $(1,1)$ given \\
\bottomrule
\end{tabular}
\end{center}

\noindent
The lower bound of constraints 5 and 6 stays at the default, because nothing
forces a given instructor-and-student pair into a tuple; the support condition of
Remark~\ref{rem:support} applies to them. It does not apply to constraints 8 and
9: constraint 10 has a lower bound of $1$, so every student is in a pair, and a
student in a pair has an instructor and a project.

\section{From Requirements to Constraints}
\label{sec:case-study}

The sections so far each added a piece. Section~\ref{sec:two-readings} showed
that a cardinality label on an edge has two opposite readings.
Section~\ref{sec:notation} replaced the label with a notation, $\Card{R}{p}{q} =
(\mathit{lower},\mathit{upper})$, that says which reading is meant, and that is no
longer limited to one constraint per edge. Section~\ref{sec:implicit} showed that
a design decides constraints it never writes down, and that a program can produce
them on demand. Section~\ref{sec:rules} gave the two rules, \emph{decomposition}
and \emph{augmentation}, that derive one constraint from another.

It remains to do the thing the introduction set out to do: take the three
business requirements and express them.

\subsection{The design}

\begin{figure}[tbp]
  \centering
  \includegraphics[width=0.55\linewidth]{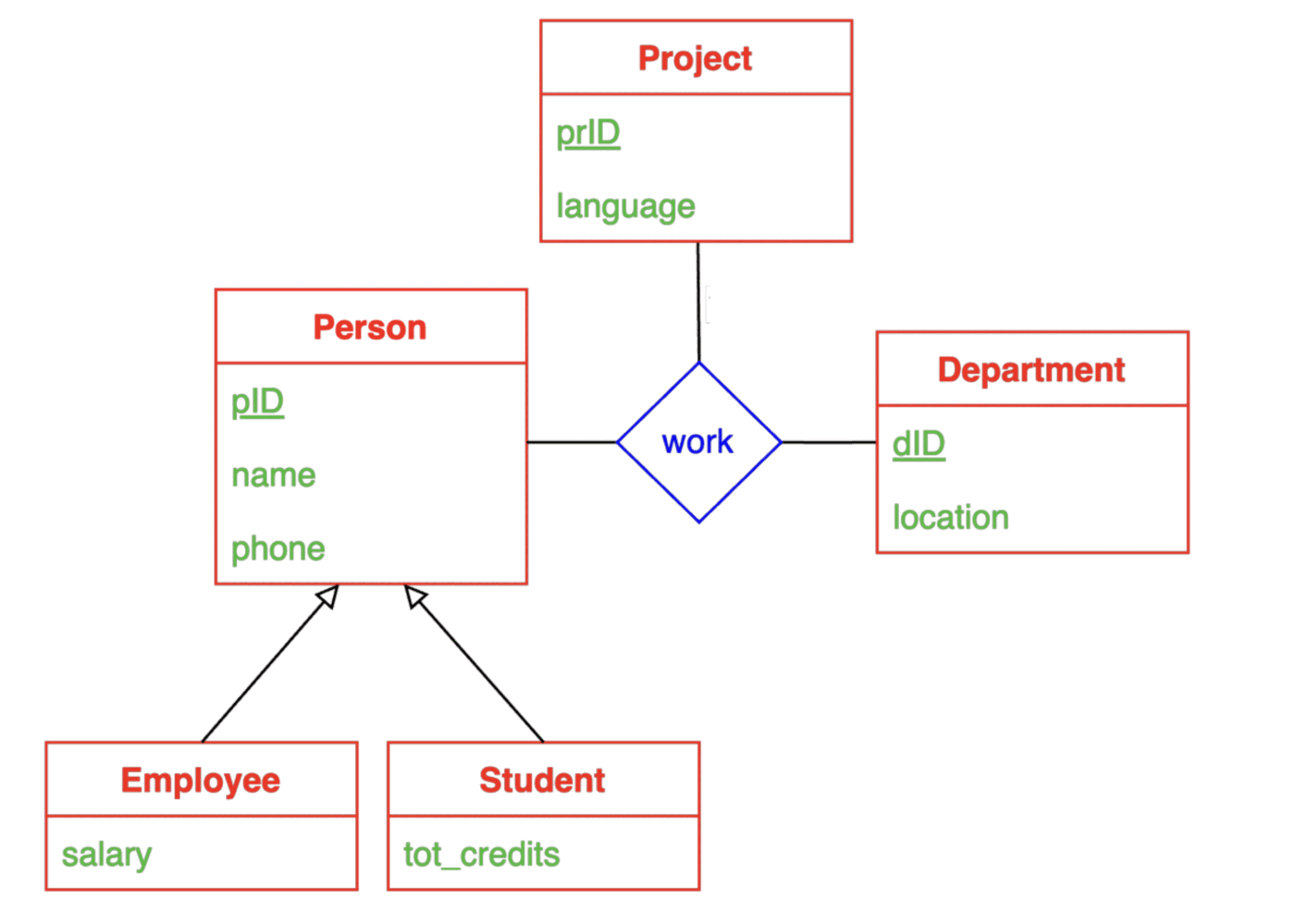}
  \caption{A sample ER design. The relationship set \work{} is ternary over
           \Pers{}, \Dept{}, and \Proj{}; \Pers{} is specialised into
           \texttt{Employee} and \Stud{}. No cardinality label appears on any
           edge, and by Section~\ref{sec:notation} there is nowhere to put the
           ones that matter.}
  \Description{An ER diagram in which Person, specialised into Employee and Student, joins Department and Project through a ternary work relationship; no edge carries a cardinality label.}
  \label{fig:research-center}
\end{figure}

The design is the university research centre of
Figure~\ref{fig:research-center}. Its entity sets are \Pers{}, \Dept{}, and
\Proj{}, and it has three requirements to satisfy.

People in the centre have a \texttt{pID}, a name, and a phone number, and come in
two kinds, \texttt{Employee} and \Stud{}; departments have a \texttt{dID} and a
location; projects have a \texttt{prID} and the language they are mostly written
in. The design has one relationship set, \work{}, and all three entity sets take
part in it.

The three requirements are:

\begin{enumerate}
  \item A person can only be assigned to at most two departments at a time.
  \item For each assigned department, a person cannot work on more than one
        project.
  \item Every project must be associated with at least one department, with at
        least one person working on it.
\end{enumerate}

\subsection{The requirements, as constraints}

Let us take the three in turn.

\paragraph{At most two departments per person.} The first requirement fixes a
person and counts departments. A person's departments are the ones appearing in
that person's \work{} tuples, and the requirement allows at most two of them. The
centre does the assigning, so a person need not be in any department. The lower
bound stays at the default of zero:

\begin{center}
$\Card{\work}{\set{\Pers}}{\set{\Dept}} = (0,2)$
\end{center}

\noindent
The requirement names the number $2$, so the upper bound is $2$. Written down,
the constraint carries the number rather than an $M$; that is, rather than an
arbitrary number.

\paragraph{At most one project per person per department.} The second requirement
counts projects per assignment, not per person. An assignment is a person
\emph{and} a department, which is what a pair-group $p$ is for. A person in two
departments can work on one project in each, so one is not a bound on
$\cmp{\set{\Pers}}{\set{\Proj}}$. The bound of one is on the pair. Nothing
requires an assigned person to work on a project in that department either, so
the lower bound is zero again:

\begin{center}
$\Card{\work}{\set{\Pers,\Dept}}{\set{\Proj}} = (0,1)$
\end{center}

\paragraph{At least one department and one person per project.} The third
requirement fixes a project and counts. ``At least one department, with at least
one person working on it'' names two conditions, one department and one person. A
count of $(\Pers,\Dept)$ pairs gives both at once: a project with a pair in it
has a department and a person, and a project with a department and a person in it
has a pair. The requirement names no upper bound, so $*$ is written there:

\begin{center}
$\Card{\work}{\set{\Proj}}{\set{\Pers,\Dept}} = (1,*)$
\end{center}

\noindent
The three requirements give three constraints:

\begin{center}
\footnotesize
\begin{tabular}{cll}
\toprule
\# & The requirement & As a constraint \\
\midrule
1 & at most two departments per person & $\Card{\work}{\set{\Pers}}{\set{\Dept}} = (0,2)$ \\
2 & at most one project per person per department & $\Card{\work}{\set{\Pers,\Dept}}{\set{\Proj}} = (0,1)$ \\
3 & at least one department and one person per project & $\Card{\work}{\set{\Proj}}{\set{\Pers,\Dept}} = (1,*)$ \\
\bottomrule
\end{tabular}
\end{center}

\subsection{The twelve constraints}

Three of the twelve are the requirements written down. The other nine follow from
the design. Writing $\cmp{p}{q}$ for $\Card{\work}{p}{q}$, the twelve constraints
are:

\begin{center}
\begin{tabular}{rlll}
\toprule
\# & Constraint & Bound & \\
\midrule
1  & \cmp{\set{\Pers}}{\set{\Dept}}          & $(0,2)$ & \explicit \\
2  & \cmp{\set{\Pers}}{\set{\Proj}}          & $(0,2)$ & \implied \\
3  & \cmp{\set{\Dept}}{\set{\Proj}}          & $(0,*)$ &\bydefault \\
4  & \cmp{\set{\Pers,\Dept}}{\set{\Proj}}    & $(0,1)$ & \explicit \\
5  & \cmp{\set{\Pers,\Proj}}{\set{\Dept}}    & $(0,2)$ & \implied \\
6  & \cmp{\set{\Dept,\Proj}}{\set{\Pers}}    & $(0,*)$ &\bydefault \\
7  & \cmp{\set{\Dept}}{\set{\Pers}}          & $(0,*)$ &\bydefault \\
8  & \cmp{\set{\Proj}}{\set{\Pers}}          & $(1,*)$ & \implied \\
9  & \cmp{\set{\Proj}}{\set{\Dept}}          & $(1,*)$ & \implied \\
10 & \cmp{\set{\Proj}}{\set{\Pers,\Dept}}    & $(1,*)$ & \explicit \\
11 & \cmp{\set{\Dept}}{\set{\Pers,\Proj}}    & $(0,*)$ &\bydefault \\
12 & \cmp{\set{\Pers}}{\set{\Dept,\Proj}}    & $(0,2)$ & \implied \\
\bottomrule
\end{tabular}
\end{center}

\noindent
Three explicit, five implied, four left at the default.

\subsection{Where the other bounds come from}

The requirements gave three of the twelve. The design decides the other nine, and
this section takes them in turn.

\paragraph{Constraint 5.} Constraint 5,
$\Card{\work}{\set{\Pers,\Proj}}{\set{\Dept}} = (0,2)$, comes from augmentation
\eqref{eq:augmentation}. A $(\Pers,\Proj)$ pair names a person, and the
departments of the pair are among the departments of that person. So a bound on
\Pers{} over \Dept{} is also a bound on the pair:

\begin{center}
$\begin{aligned}
  &\Card{\work}{\set{\Pers}}{\set{\Dept}} = (0,2)\\
  \Longrightarrow\;&\Card{\work}{\set{\Pers,\Proj}}{\set{\Dept}} = (0,2)
\end{aligned}$
\end{center}

\noindent
Both ends of the bound are reachable. A pair that is in no tuple counts zero, and
a pair whose person works on one project in two departments counts two. So the
bound cannot be any tighter than $(0,2)$. The requirements say nothing about
$(\Pers,\Proj)$ pairs.

\paragraph{Constraints 8 and 9.} These come from constraint 10. Constraint 10
counts $(\Pers,\Dept)$ pairs per project. Constraint 8 counts persons per
project, and constraint 9 counts departments per project. Every tuple of a
project names one person and one department, so neither count can be zero. That
is where the lower end of $1$ comes from. Neither count has an upper end: no
requirement names one, and a project can have as many people and as many
departments as we like.

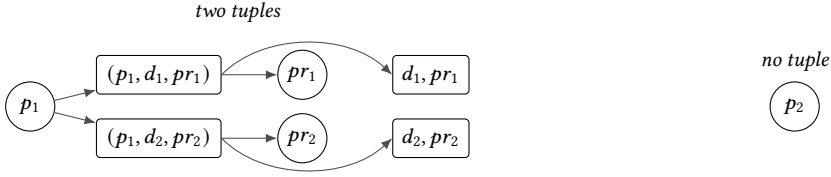
\begin{figure}[tbp]
\centering
\begin{minipage}[c]{0.47\linewidth}\centering
  {\footnotesize\itshape two tuples}\\[1mm]
  \begin{tikzpicture}[x=1mm,y=1mm]
    \node[ent] (g)  at ( 0,  0)  {$p_1$};
    \node[tup] (t1) at (17, 4.2) {$(p_1,d_1,pr_1)$};
    \node[tup] (t2) at (17,-4.2) {$(p_1,d_2,pr_2)$};
    \node[ent] (c1) at (36, 4.2) {$pr_1$};
    \node[ent] (c2) at (36,-4.2) {$pr_2$};
    \node[tup] (e1) at (53, 4.2) {$d_1, pr_1$};
    \node[tup] (e2) at (53,-4.2) {$d_2, pr_2$};
    \draw[ar] (g)--(t1);  \draw[ar] (g)--(t2);
    \draw[ar] (t1)--(c1); \draw[ar] (t2)--(c2);
    \draw[ar] (t1.east) .. controls +(5.5,5.6) and +(-6,5.6) .. (e1.west);
    \draw[ar] (t2.east) .. controls +(5.5,-5.6) and +(-6,-5.6) .. (e2.west);
  \end{tikzpicture}
\end{minipage}\hfill
\begin{minipage}[c]{0.47\linewidth}\centering
  {\footnotesize\itshape no tuple}\\[1mm]
  \begin{tikzpicture}[x=1mm,y=1mm]
    \node[ent] (g) at (0,0) {$p_2$};
  \end{tikzpicture}
\end{minipage}
\caption{Why constraints 2 and 12 are $(0,2)$. A person is in at most two tuples,
         so they name at most two projects and at most two
         $(\Dept,\Proj)$ pairs; a person in no tuple at all counts zero.}
\Description{Two graphs side by side. The left shows one Person joined through two tuple boxes, each naming a distinct Project and a distinct Department-Project pair; the right shows a Person node in no tuple.}
\label{fig:two-tuples}
\end{figure}

\paragraph{Constraints 2 and 12.} These come from constraints 1 and 4 together.
Constraint 4 allows one project per $(\Pers,\Dept)$, and constraint 1 allows two
departments per person. A person is therefore in at most two tuples. Constraint 2
counts projects per person, and each tuple names one project, so that count is at
most two. Constraint 12 counts $(\Dept,\Proj)$ pairs per person, and two tuples
in the same department name the same project, so that count is at most two as
well. The lower end of both is $0$, because constraint 1's lower end is $0$: a
person need not be in any department, and a person in no tuple counts nothing.
Figure~\ref{fig:two-tuples} draws both cases.

\paragraph{The remaining four.} Constraints 3, 7, and 11 stay at the default.
Constraint 3 counts from a department; 7 and 11 likewise, and 6 counts from a
$(\Dept,\Proj)$ pair. No requirement forces a department to appear, so the lower
end is $0$ in all four. No requirement caps those counts either, so the upper end
is $*$: a department can carry any number of people, projects, or
$(\Pers,\Proj)$ pairs, and one $(\Dept,\Proj)$ pair can carry any number of
people.

\section{Related Work}
\label{sec:related}

The two-reading problem is a consequence of the notation rather than of any
particular theory, and it has been visible since the beginning.
\citet{chenEntityrelationshipModelUnified1976} introduced the ER model with
cardinality on the edges, and the reading his notation induces is the one we call
\emph{look here}. The opposite reading arises from the translation to relational
schemas and is the one taught by standard textbooks such as
\citet{silberschatzDatabaseSystemConcepts2020}. \citet{oliveConceptualModelingInformation2007}
gives a systematic treatment of cardinality constraints for conceptual
schemas, and in Chapter~4 of the constraint types themselves; our notation is
intended to be readable in his setting as well as in the graphical one.

For $n$-ary relationships specifically, \citet{mcallisterCompleteRulesNary1998}
is the closest antecedent to the present work. McAllister unifies the
relationship cardinality constraints of several existing approaches into a common
framework, introduces a tabular notation for specifying them, and gives rules for
detecting inconsistencies among constraints specified for one relationship. Our
constraint enumeration in Section~\ref{sec:notation} arrives at the same class of
objects for the ternary case, and our decomposition and augmentation rules play
the role his consistency rules play, but we are concerned with a different
question: which of the constraints a design determines \emph{without} specifying
them, and what a derived bound can soundly be. \citet{cuadraGuidelinesRepresentingComplex2013}
give practical guidelines for detecting and representing cardinality constraints
in ternary relationships. They motivate the same difficulty we begin from,
namely that designers do not always know how to manage the constraints of a
ternary relationship, but they do not propose a notation free of the reading
ambiguity.

The generalisation of cardinality constraints to arbitrary groups of entity sets,
and the associated implication problem, has been studied extensively.
\citet{hartmannImplicationProblemCardinality2001} characterises closed sets of
cardinality constraints, gives a finite axiomatisation for their implication
problem, and constructs Armstrong databases; \citet{linkCardinalityConstraintsFunctional2023}
extend the class to cardinality constraints together with functional
dependencies over SQL tables and give a quadratic-time implication test.
\citet{ReasoningParticipationConstraints} reasons about participation
constraints (the lower bounds, in our terms) and about the constraints Chen's
model induces, which is precisely the boundary our default of $0$ sits on. Our
contribution is deliberately narrower than that literature: we do not
characterise the implication problem, we give two sound rules that can be applied
without a solver, and we are explicit about the asymmetry
(Remark~\ref{rem:support}) that a purely equational statement of augmentation
hides.

Finally, the lossiness of decomposing an $n$-ary relationship into binary ones
is the standard reason a designer cannot simply sidestep the notation.
\citet{jonesTernaryRelationshipDecomposition1996} identify the conditions under
which the decomposition is lossless and dependency-preserving, and show that they
do not always hold, which is why we treat splitting as one of the two
workarounds that cost more than they appear to.

\section{Conclusion}
\label{sec:conclusion}

A cardinality label on the edge of an ER diagram is an ambiguous object. It has
two readings, they are exact opposites, and for relationships of three or more
entity sets the diagram is not merely ambiguous but inexpressive: a ternary
relationship set admits twelve cardinality constraints and the diagram can carry
three of them.

The notation $\Card{R}{p}{q} = (\mathit{lower},\mathit{upper})$ resolves both
problems at once. It names the group being iterated over and the group being
counted, so it says what it means without taking a side in the look-here /
look-across debate, and because it is a constraint rather than a mark on a
picture, a relationship set can carry as many of them as it has. Writing them
down as data is also what makes them something a program can read, and the
constraints nobody wrote down have a meaning too: the default is $(0,*)$, and a
design determines some of the rest on its own.

Two rules carry most of that determination. Decomposition transfers an upper
bound from a pair-group down to a subgroup; augmentation transfers an upper bound
from a group to a finer group on the other side, and not the matching lower one
(Remark~\ref{rem:support}). Both lose information, and neither produces a bound
tighter than the default on its own authority. They do not account for
everything: constraints 2 and 12 of the case study are neither, and fall out of
counting the tuples a single person can be in.

The case study of Section~\ref{sec:case-study} is the shape of the intended use.
Three business requirements become three explicit constraints; two rules, a
default, and a counting argument over the tuples of a single person yield nine
more; and the designer is left reviewing nine statements the design had already
made without saying so, at which point a loose bound, or a contradictory pair,
is something to be told about rather than something to be discovered in
production.

\bibliographystyle{ACM-Reference-Format}
\renewcommand{\bibfont}{\small}
\bibliography{refs}

\end{document}